\documentclass{iaus}
\usepackage{graphicx}

\title[Stellar archeology of nearby LINERs] 
{Stellar archeology of the nearby \\
LINER galaxies NGC 4579 and NGC 4736
}

\author[J.E. Steiner et al.]   
{J.E. Steiner$^{1\dagger}$, R.B. Menezes$^1$, T.V. Ricci$^1$ \and A.S. Oliveira$^2$}

\affiliation{$^1$Instituto de Astronomia, Geof\'isica e Ci\^encias Atmosf\'ericas, Universidade de S\~ao Paulo, \\
Rua do Mat\~ao, 1226, S\~ao Paulo - SP, Brasil \\[\affilskip]
$^2$IPD/Univap \\ $^{\dagger}${\tt steiner@astro.iag.usp.br}}

\pubyear{2009}
\volume{262}  
\pagerange{1--2}
\jname{Stellar Populations - Planning for the Next Decade}
\editors{Gustavo R. Bruzual \& Stephane Charlot}
\begin{document}

\maketitle

\begin{abstract}

Stellar archeology of nearby LINER galaxies may reveal if there is a stellar young population that may be responsible for the LINER phenomenon. We show results for the classical LINER galaxies NGC 4579 and NGC 4736 and find no evidence of such populations.
\end{abstract}

\section{Introduction}

Low Ionization Nuclear Emitting Regions - LINERs - are a subset of Active Galactic Nuclei - AGN- quite common in early type galaxies. A variety of models have been proposed to explain their properties: shocks, photoionization by low luminosity AGNs, young or old stars etc. Some controversy still persists on the mechanism of ionization of the narrow line region - NLR. Our goal is to study nearby galaxies in great detail to identify which mechanism or mechanisms are responsible for the LINER emission. Therefore it is important to understand the stellar population history in order to search for possible ionizing populations. 

\section{Observations and Methodology}

We used high spatial resolution integral field spectroscopy from the IFU-GMOS \\ \cite[(Allington-Smith et al. 2002)]{Allington-Smith et al.(2002)} on the Gemini North telescope to investigate the stellar populations of the LINER galaxies NGC 4579 and NGC 4736, aiming to decipher its star-formation and chemical histories. After correction for differential atmospheric refraction and Richardson-Lucy de-convolution of the data cube, we synthesize the stellar population using the software {\sc starlight} (\cite[Cid Fernandes et al. 2005]{Cid Fernandes et al. 2005}) and present it in form of images of distinct metallicities, ages and kinematical properties. A paper with full set of results will be published elsewhere.

\section{Results and conclusions}

\begin{figure}[t]
\begin{center}
\begin{minipage}[b]{0.45\linewidth}
\includegraphics[width=1.0 \columnwidth,angle=0]{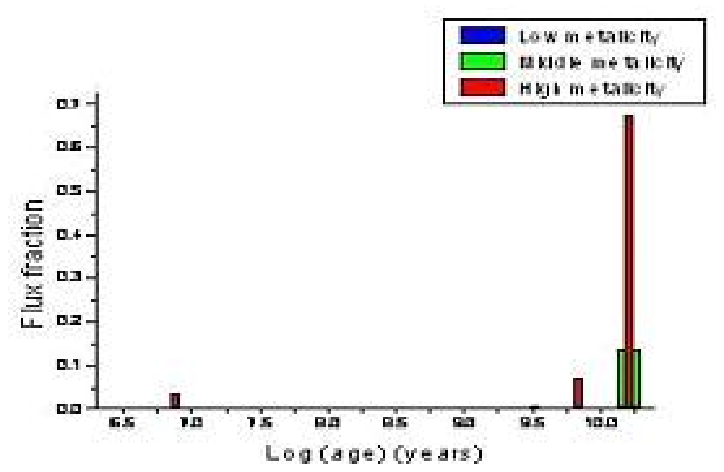}
\end{minipage} \hfill
\begin{minipage}[b]{0.45\linewidth}
\includegraphics[width=1.0 \columnwidth,angle=0]{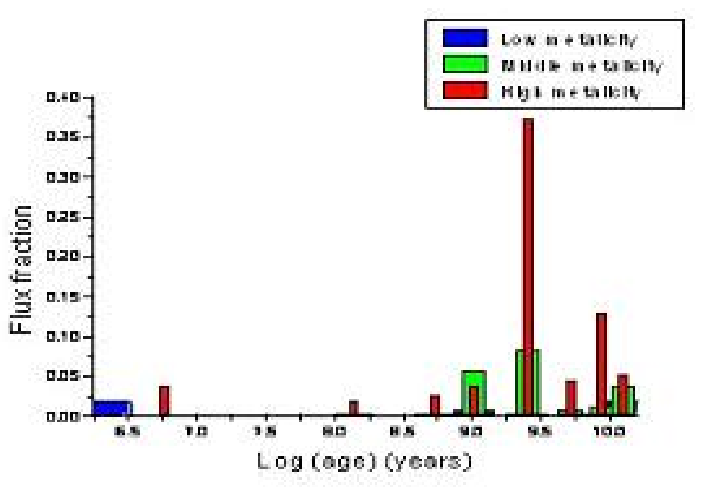}
\end{minipage}
\caption{Star formation and chemical histories of NGC 4579 (left) and NGC 4736 (right).}
\label{fig1}
\end{center}
\end{figure}

\begin{figure}[h]
\begin{center}
\begin{minipage}[b]{0.45\linewidth}
\includegraphics[width=1.0 \columnwidth,angle=0]{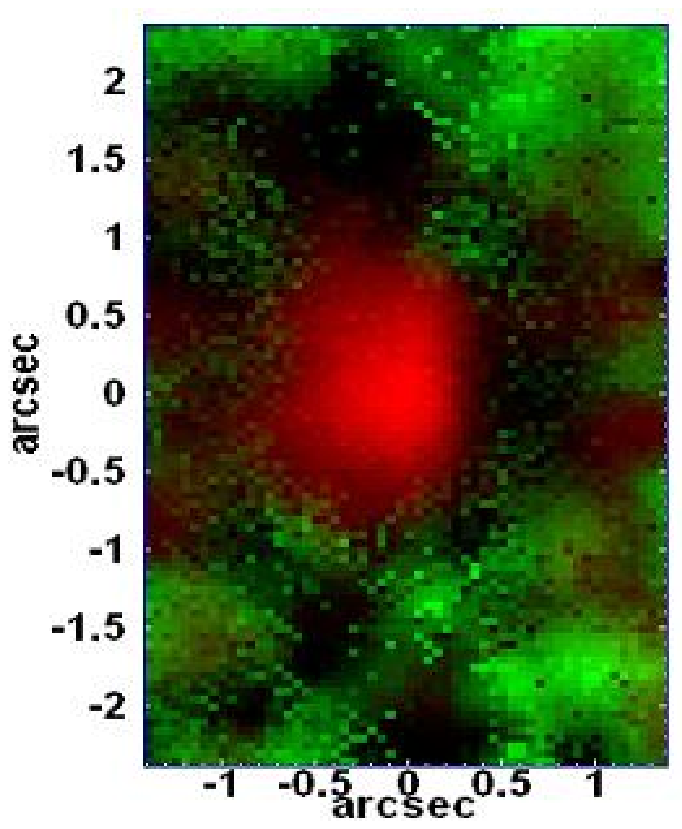}
\end{minipage} \hfill
\begin{minipage}[b]{0.45\linewidth}
\includegraphics[width=1.0 \columnwidth,angle=0]{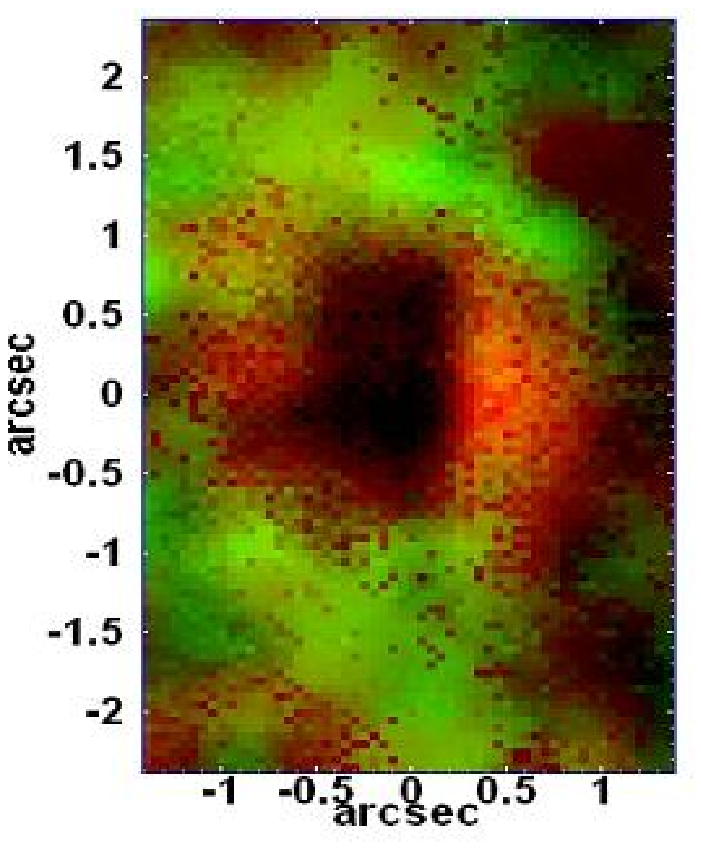}
\end{minipage}
\caption{The images of light fraction of the NGC 4736 central bulge.  
\textbf{Left}: the first generation of stars. Notice that the bulge is centered at (-0.1''; 0.0'') while the type 1 AGN is at (x=0.3; y= 0.1) as shown by \cite[Steiner et al. (2009)]{Steiner et al.(2009)}. 
\textbf{Right}: the second generation of stars.}
\label{fig2}
\end{center}
\end{figure}

	In NGC 4579 the star formation history is quite simple (Fig. \ref{fig1}). Nearly 100\% of the stellar mass has an age of $\sim$ 9-12 Gyr, although some recent star formation, of negligible mass fraction, is also detected. The first generation has two metallicities: metal rich stars are highly concentrated at the center while metal poor stars form an annulus around it. A small fraction of the stellar mass formed an external ring and is metal enhanced.
	In NGC 4736 we found that most of the stars were formed at two epochs: nearly 1/3 of the light comes from old stars (T $>$ 10 Gyr) (Fig \ref{fig1}). The metal rich ones (about solar) are highly concentrated in the nucleus while metal poor stars are scattered around it. Nearly 2/3 of the light is emitted by stars formed about 2.5 Gyr ago. These stars come in two metallicities: metal enriched (2.5 solar) form an oval structure around the nucleus while metal poor (0.21 solar) form a bar nearly aligned with the axis of rotation (Fig. \ref{fig2}). A smaller fraction of the stars were formed more recently.  
	Our findings illustrate some of the processes of classical bulge and pseudobulge formation. Although there are a variety of populations in term of ages, metallicities and structures, we do not see any structure that could somehow be associated to the LINER ionization characteristic.

\end{document}